# FRAGMENTATION OF RELATIVISTIC OXYGEN NUCLEI IN INTERACTIONS WITH A PROTON

**V.V.Glagolev[1], K.G.Gulamov, V.D.Lipin, S.L.Lutpullaev, K.Olimov, Kh.K.Olimov, A.A.Yuldashev and B.S.Yuldashev[2]**

Physical-technical institute of SPA "Physics—Sun" of Uzbek Academy of Sciences, G.Mavlyanova str., 2B, Tashkent, 700084, Uzbekistan

## Abstract

The data on investigation of inelastic interactions of $^{16}$O nuclei with a proton at 3.25 A GeV/c momentum by the bubble chamber method are presented. The separate characteristics as fragments isotopic composition and as topo-logical cross sections of fragmentation channels are given. The processes of light fragments formation and breakup of $^{16}$O nucleus on multicharge fragments have been investigated. The comparison of experimental data with the calculations by statistical multifragmentation model was conducted.

## Introduction

The fragmentation of excited atomic nuclei formed in hadron- and hadron-nuclear interactions is the one of the fundamental problems in nuclear physics. The experimental data show that, at energies about some Gev per nucleon it gives a principal contribution to multiplicity of the secondary particles [1,2]. A number of experimental and theoretical works have been devoted to the investigation of this phenomenon. But there is no full comprehension of this phenomenon now, and the existing model calculations are able to depict only some details of that [3-5]. It makes the further investigations and collecting of experimental data with identification of all secondary particles and nuclei with a possibility of study of fragments yield correlation in this field to be important. These correlations, being sensitive to fragmentation processes

---

[1] Joint Institute for Nuclear Research, Dubna, Moscow region, 141980 Russia.
[2] Institute of Nuclear Physics, Uzbek Academy of Sciences, pos. Ulughbek, Tashkent, 702132 Republic of Uzbekistan.

mechanisms, may be a critical test for checking of different model approximations.

The new experimental results on fragmentation of oxygen nucleus in interactions with a proton at 3.25 A GeV/c momentum are presented. These data can be usefull in solution of astrophysical problems, because in interstar space, consisting basicly of hydrogen atoms; cosmic ray composition is defined mostly by nuclei fragmentation prosecces in interactions with hydrogen.

The experimental results were systematically compared with the predictions of SMM (statistical multifragmentation model) designed for the statistical modeling of the nucleus breakup in interactions with proton [6]. The fermi-breakup of the residual excited nucleus after in-nuclear cascade is the main mechanism of fragments formation at SMM calculations [7]. The contribution of the $^5$He, $^5$Li, $^8$Be and $^9$B unstable nuclei decays to the final state was also taken into account in model. For the aims of SMM calculations more than 22000 events were modeled in this paper.

### 1.The method of the experiment

The 1-meter hydrogen bubble chamber of JINR (Dubna, Russia), installed in the magnetic field, was used by us in this experiment. The chamber was exposed with the oxygen ($^{16}$O) nuclei beams at the 3.25 A GeV/c momentum. The homogeneity of the target and the low density of the working liquid allowed us to identify with high confidence all of the secondary particles and fragments charges, and to measure with high precision their momenta. The space reconstruction and kinematical analysis of measured events were being conducted with use of adapted library programmes **CERN-HIDRA[8]**. The methodical questions connected with treatment of film information and identification of secondary particles and nuclei are shown in papers **[9-11]**. For an analysis 11000 events were selected.

The mean relative errors in the measurements of the surely identified protons (mainly the recoil protons) momenta and those of $\pi^{\pm}$ mesons without any restrictions on the tracks lengths proved to be $<\Delta P/P>=(4.56\pm 0.01)\%$ and $<\Delta P/P>=(2.65\pm0.05)\%$ respectively.

The features of the secondary particles, formed in the $^{16}Op$ interactions were obtained at their lengths L>40 cm in the working volume of the chamber. The corrections, connected with losses caused by the secondary interactions at the length up to 40 cm., were included at the physical analysis of the data. In such a selection, the mean relative error at the designation of the momenta did not exceed of 3.5 % in all values of charges. The absolute errors at the measurements of the azimuthal angle in the XOY plane proved to be $<\Delta\beta>=(0.60\pm0.01)$ mrad and $<\Delta\alpha>=(1.5\pm0.02)$ mrad for the depth angle.

In such errors value of momentum measurement, fragments with $z_f\leq4$ are well separated by mass.

The full cross section of $^{16}Op$ interactions at 3.25 A GeV/c proved to be (395±10) mb, herewith (334±6) mb accounted for inelastic channels.

## 2. The isotopic composition of the secondary fragments, and the topological cross sections of the fragmentation channels

In this chapter, the data on the isotopic composition of the secondary particles, formed in the oxygen nucleus fragmentation processes in the $^{16}Op$ interactions at the 3.25 A GeV/c momentum, are presented. The part of the fragments with the $Z_f$ charge was being defined by the analysis of distributions on the X= 1/p variable, where p-was the fragment momentum in laboratory frame. Herewith, the fragments with a measured length L≥35 sm for $z_f\leq2$ and L≥40 sm for $z_f\geq3$ were selected.

The multinucleon fragments spectra by X for given fragments charge are well depicted by sum of gauss distributions corresponding to each isotope. But the proton spectrum part for one-charge particles we managed to describe well by sum of two gauss functions with distinct gauss distributions width when restricting ejection angle θ<3.6 degree. Such a restriction by ejection angle is corresponding to protons with a transverse momentum ≤200 MeV/c. The distribution of one-charge fragments spectrum with such a restriction is presented in fig.1, and obtained parts of isotopes are shown in tab.1. In the same table the results obtained for fragments with $z_f=2\div8$ and calculated by SMM data with above-mentioned restriction for one-charge fragments are presented either. In tab.1 $\Delta W$ is an error of defining isotopes part.

**Table 1**

The part of the different isotopes formed in the $O^{16}p$ interactions (%)

| Z | A | Experiment | | SMM | |
|---|---|---|---|---|---|
| | | W | $\Delta W$ | W | $\Delta W$ |
| 1 | $^1H$ | 67.7 | | 64.3 | |
| | $^2H$ | 22.9 | 0.6 | 24.6 | 0.4 |
| | $^3H$ | 9.4 | | 11.1 | |
| 2 | $^3H$ | 19.8 | | 29.0 | |
| | $^4H$ | 79.7 | 0.9 | 68.5 | 0.5 |
| | $^6H$ | <0.5 | | 2.5 | |
| 3 | $^6Li$ | 55.1 | | 46.1 | |
| | $^7Li$ | 32.4 | 2.2 | 36.9 | 1.3 |
| | $^8Li$ | 12.5 | | 17.0 | |
| 4 | $^7Be$ | 56.6 | | 55.7 | |
| | $^9Be$ | 38.3 | 3.0 | 16.4 | 1.2 |
| | $^{10}B$ | 5.1 | | 27.9 | |
| 5 | $^{10}B$ | 48.2 | | 42.5 | |
| | $^{11}B$ | 49.5 | 1.9 | 50.2 | 1.1 |
| | $^{12}B$ | 2.3 | | 7.3 | |
| 6 | $^{10}C$ | 3.5 | | 16.2 | |
| | $^{11}C$ | 18.2 | | 38.4 | |
| | $^{12}C$ | 52.2 | 1.5 | 19.8 | 1.1 |
| | $^{13}C$ | 18.8 | | 20.1 | |
| | $^{14}C$ | 7.3 | | 5.5 | |
| 7 | $^{13}N$ | 9.4 | | 8.9 | |
| | $^{14}N$ | 42.1 | 1.2 | 47.7 | 0.8 |
| | $^{15}N$ | 48.5 | | 43.4 | |
| 8 | $^{14}O$ | 6.1 | | 3.3 | |
| | $^{15}O$ | 66.2 | 1.5 | 93.5 | 0.9 |
| | $^{16}O$ | 27.7 | | 3.2 | |

Let us see tab.2 where inelastic cross-sections of topological channels of fragmentation are presented. The charge composition of fragments is pointed in brackets, and corresponding cross sections for experimental and model data are shown below. The Z=1 topology corresponds to channels without formation of multicharge fragments. It is seen from from tab.2 that, cross section of two-charge fragments yield accounts for a significant part of inelastic cross section.

It is interesting to mention an absence in experimental data of (44), (35), (233) topological channels where the total charge is equal to that of initial nucleus.

**Table 2**

Inelastic cross sections of the oxygen nuclei fragmentation (mb)

| Topology | ($z_f$=1) | (2) | (22) | (222) |
|---|---|---|---|---|
| Experiment | 6.43 ± 0.46 | 23.58 ± 0.88 | 36.44 ± 1.10 | 31.27 ± 1.02 |
| SMM | 5.19 ± 0.28 | 15.03 ± 0.48 | 19.69 ± 0.55 | 12 .42 ± 0.43 |
| Topology | (2222) | (23) | (24) | (25) |
| Experiment | 3.51 ± 0.34 | 11.53 ± 0.62 | 6.66 ± 0.47 | 7.45 ± 0.50 |
| SMM | 0.29 ± 0.07 | 13.83 ± 0.46 | 17.35 ± 0.51 | 20.02 ± 0.50 |
| Topology | (26) | (223) | (224) | (233) |
| Experiment | 10.14 ± 0.58 | 3.11 ± 0.32 | 0.93 ± 0.18 | _ |
| SMM | 23.32 ± 0.59 | 3.35 ± 0.23 | 1.06 ± 0.13 | 0.05 ± 0.03 |
| Topology | (3) | (33) | (34) | (35) |
| Experiment | 5.29 ± 0.41 | 1.26 ± 0.20 | 0.66 ± 0.15 | _ |
| SMM | 12 .49 ± 0.44 | 2.50 ± 0.19 | 2.81 ± 0.21 | 0.15 ± 0.06 |
| Topology | (4) | (44) | (5) | (6) |
| Experiment | 5.60 ± 0.43 | _ | 16.46 ± 0.73 | 54.16 ± 1.34 |
| SMM | 13.09 ± 0.45 | 0.71 ± 0.11 | 21.39 ± 0.57 | 47.54 ± 0.85 |
| Topology | (7) | (8) | | |
| Experiment | 65.35 ± 1.47 | 46.70 ±1.24 | | |
| SMM | 75.51 ± 1.01 | 29.16 ±0.67 | | |

In (34±1)% cases at final state two and more multicharge fragments are observed which data are close to those calculated by SMM (36±1)%. But in experiment the cross section of observance of two and three helium nuclei twice as much as that predicted by model, and this one of four two-charge fragments formation is significantly deviating. Let us point out also, that a probability of two-charge fragment accompaning by heavier fragment is substantially greater in model than in experiment.

### 3.Formation of the light ($H^1$, $H^2$, $H^3$, $He^3$) fragments

In this chapter the results of study of mechanisms of light fragments formation with masses $A_f \leq 3$ are presented.

For the final identification the separation of the fragments implemented by their momenta: the one-charge fragments with P=(4.6÷7.8) GeV/c were taken as $^2$H nuclei and those with P>7.8 GeV/c as $^3$H nuclei, and two-charge ones with P<10.8 GeV/c as $^4$He nuclei. The one-charge positive relativistic particles with 1.5<P<4.6 GeV/c were referred to protons.

The energy and angular features had been studied in antilaboratory frame, i.e. in projectile-nucleus rest system. In this frame the kinetic energy of nuclei with A=2 and 3 does not exceed 200 MeV. Further the same border was settled up for protons referred to fragments. In such a selection we might neglect mixture of $\pi^+$ mesons among positive one-charge particles referred to protons.

### 3.1.The mean multiplicities and inclusive cross sections

In tab.3 the mean multiplicities <$n_f$> and inclusive cross sections $\sigma_{in}$ of the light $^1$H, $^2$H, $^3$H, and $^3$He fragments yields are presented.

We must point that, as it is seen from tab.3, the cross sections of the «mirror» $^3$H and $^3$He nuclei formation in the statistical errors range are coinciding. The mean associative multiplicities of the multinucleon fragments with the yield of the one $^3$He or $^4$He nucleus have been also considered in semi-inclusive reactions. These data are presented in tab.4.

It is seen from tab.4 that, the mean multiplicities of the associated particles appeared to be the same for both isoduplicates also. The experimental results point, that the angular spectrums are also coinciding for these nuclei: <$\theta(^3$He)>=(1.43±0.03)$^0$, $\sigma(\theta(^3$He))=1.03$^0$, and <$\theta(^3$H)>= (1.42±0.04)$^0$, $\sigma(\theta(^3$H))=1.02$^0$.

The similar effects we may probably expect in reactions of projectile-nucleus breakup on fragments with an absence of nuclei with mass $A_f>3$. Such events may be referred to reactions of oxygen nucleus full breakup, because there are no fragments, which could be formed from $\alpha$-clusters among them. The total cross section of these channels proved to be $15.54\pm 0.78$ mb.

**Table 3**

The mean multiplicities $<n_f>$ and the inclusive cross sections $\sigma_{in}$ of the light $H^1$, $H^2$, $H^3$, and $He^4$ fragments yields

| Fragment Type | $^1$H | $^2$H | $^3$H | $^3$He |
|---|---|---|---|---|
| $<n_f>$ | 1.525±0.017 | 0.350± .004 | 0.125±0.001 | 0.126± .001 |
| $\sigma_{in}$, mb. | 549.0 ± 6.1 | 126.0 ±1.4 | 45.0 ±0.4 | 45.4 ±0.4 |

**Table 4**

The associative multiplicities of light fragments in the channels with the «mirror» $^3$H and $^3$He nuclei formation

| Type of "mirror" nucleus | $^3$H | $^3$He |
|---|---|---|
| Particle type | | |
| $^2$H | 071 ± 0.02 | 0.71 ± 0.02 |
| $^3$H | 1.0 | 0.29 0.01 |
| $^3$He | 0.28 ± 0.01 | 1.0 |
| $^4$He | 0.85 ± 0.03 | 0.86 ± 0.03 |
| A(z≥3) | 0.25 ± 0.02 | 0.21 ± 0.02 |

It is seen from tab.5 that, upon the oxygen nucleus full breakup on fragments with mass number $A_f\leq3$, the mean multiplicities of the «mirror» $^3$H, and $^3$He nuclei appeared to be as close as in the inclusive $^{16}$Op-reactions. When comparing the features of the full breakup channels with the absence of the one of isoduplicates with $A_f=3$ (n($^3$He)=0, and n($^3$H)=0 channels), we derived, that not only the mean multiplicities of $^3$H and $^3$He nuclei, but also the realization probabilities of those were coinciding. The mean multiplicities of the $^2$H nuclei are also the same in these channels. Such an isotopic symmetry takes also place in the events with yields of both «mirror» nuclei (n($^3$H)$\geq$ 1, n($^3$He$\geq$1)).

**Table 5**

The mean multiplicities of the light fragments, and $\pi$-mesons in the different channels of the oxygen nucleus full breakup

| Channels | Part of events (%) | <n($^2$H)> | <n($^3$H)> | <n($^3$He)> | <n($\pi^-$)> |
|---|---|---|---|---|---|
| n($^3$He)=0 | 44.2 ± 3.4 | 1.76 ± 0.13 | 1.08 ± 0.08 | _ | 0.43 ± 0.03 |
| A$_f$ ≤3 | 100 | 1.40 ± 0.07 | 0.82 ± 0.04 | 0.83 ± 0.04 | 0.50 ± 0.03 |
| n($^3$H)=0 | 41.4 ± 3.3 | 1.70 ± 0.13 | _ | 1.11 ± 0.09 | 0.58 ± 0.05 |
| n($^3$H)≥1 n($^3$He) ≥1 | 29.0 ± 2.7 | 0.84 ± 0.08 | 1.25 ± 0.12 | 1.27 ± 0.12 | 0.58 ± 0.05 |
| n($^3$H) =0 n($^3$He) ≥1 | 28.5 ± 2.7 | 1.49 ± 0.14 | _ | 1.61 ± 0.15 | 0.53 ± 0.05 |
| n($^3$H) ≥1 n($^3$He)=0 | 29.6 ± 2.8 | 1.59 ± 0.15 | 1.55 ± 0.14 | _ | 0.32 ± 0.03 |

It is also seen from these data that, the $\pi$-mesons mean multiplicity is connected with the charge of the fragments with A$_f$=3. It is substantially more in the events with $^3$He formation than in those with $^3$H. Because the process of the inelastic recharging of the oxygen nucleus neutron n→p+$\pi^-$ is the main source of the $\pi$-mesons formation at the energies near of 1 GeV, the residual compound nucleus in the events with the relatively rapid $\pi$-meson formation will be, in average, proton excessive, and thus may cause the increasing of the $^3$He yield probability.

The observed behavior of isoduplicates cross sections points probably that the nucleus coulomb forces do not affect substantially the fragment formation, and the probability of supplimentary charge transfering to residual excited nucleus is very small.

The inclusive cross sections of the «mirror» nuclei formation with A=7 ($^7$Li, $^7$Be) were also defined. They are also coinciding within the statistical errors: $\sigma_{incl}(^7$Li$)=9.3\pm0.5$ mb, $\sigma_{incl}(^7$Be$)=9.1\pm0.5$ mb.

## 3.2. The energy and angular spectrums

Let us see the inclusive and energy spectra of $^2$H, $^2$H, $^3$H, and $^3$He fragments in antilaboratory frame. The invariant structural function f(T) = Ed$^3\sigma$/d$\mathbf{p}^3$ of protons versus to their kinetic energies T dependence is presented in fig.2, the results by the SMM calculations are also given there. It is seen from the figure that, the experimental spectrum is divided into two regions characterized by the different slopes. In the region with T<20 MeV the spectrum decreases speedy as the energy increases, and in that with T>20 MeV the spectrum becomes more flat approaching to the exponential one. The model calculations, on the whole, show the same behaviour. Herewith, in the T>50 MeV region, where the products of in-nuclear cascade have to dominate, the model satisfactorily depicts the experimental data. Fig.2b shows it more clearly. The ratio of the experimental invariant structural function to the theoretical one is presented in this figure. It is seen from this figure that, the model decreases substantially the cross section of the low energy protons. There are also deviations in transitive 20<T<50 MeV region. As in the T<20 MeV kinematical region the decisive contribution is given by the Fermi-breakup, the obtained result probably points that, the contribution of the «evaporation» processes (before the equilibrium stage) should be taken into account in the SMM model.

Let us see further the experimental data for the $^2$H, $^3$H, $^3$He-multinucleon fragments. The distribution of the invariant structure function F(T) in experiment and that by the model calculations are given in fig.3,a,b,c. As it is seen from figure, the experimental spectra of $^2$H, $^3$H, and $^3$He in comparison with those of proton show more flat slope, but also are divided into two regions: T≤20 MeV, and T>25 MeV. The model data behavior is different from that of the experiment. Herewith, as it is seen from fig.3, in the low energy region T≤20 MeV the SMM agrees satisfactorily with the experiment. As to the rapid fragments (T>25 MeV), there are deviations between model and experiment which amplify as the kinetic energy increases. This fact, probably, may be connected with the presence of the supplementary processes of the rapid fragments formation not counted in the model. In this kinematical region, as it is known, the coalescence process of the cascade nucleons with the little relative momenta may be such a supplementary mechanism [12].

As it is known, the cross section of fragments formation with mass number $A_f$ is expressed by neglecting of both proton and neutron spectrums difference through the protons yield cross section as following:

$$E_A d^3\sigma_A/d\mathbf{p}^3_A = C_A(E_p d^3\sigma_p/d\mathbf{p}^3_p)^A \qquad (1)$$

Where $P_p$- proton momentum, $P_A=A_f*P_p$, $C_A$-coefficient of coalescence [13]. The existing data [7] point that, $C_A$ depends slightly on the target mass, and does not depend on both projectile energy and fragment ejection angle. The results of the corresponding calculations are shown by the solid lines in fig.3, herewith the proton spectrum was used from our experiment. $C_A$ coefficient was being defined by the extrapolation to the corresponding experimental data in the T≥75 MeV region. It is seen from fig.3 that, the light fragments spectrums are in accordance with the coalescence model predictions in the T>70 MeV region. Comparing the experimental spectrums of the $^2$H, $^3$H, $^3$He fragments with those of both SMM and coalescence models, we may probably make conclusion that there are contributions of two light nuclei formation mechanisms at least. One of them is the process of residual thermal nucleus Fermi-breakup. The other is the coalescence process of the rapid nucleons formed at in nuclear cascade.

As it was mentioned above, the data of this experiment point that, the physical conditions of the mirror $^3$H and $^3$He nuclei formation in the proton interactions with the double magic $^{16}$O-oxygen nucleus are the same. Besides the inclusive cross sections, the angular spectra, the mean multiplicities of the associated $^2$H nuclei and $^4$He ones are also coinciding. The identity of the invariant cross sections distributions of the «mirror» nuclei with A=3, as it is seen from fig.3, confirms this inference.

Let us see the angular distributions of the little energy (T<10 MeV per nucleon) $^1$H, $^2$H, $^3$H, $^3$He fragments, which are presented in fig.4 (taking into account the identity of the angular features of $^3$H and $^3$He nuclei, we associated with corresponding data in fig.4). These distributions, in whole, are not isotropic and have the little asymmetry to forward direction. In the same place, the results of the SMM calculations are presented (the solid lines). The calculated data are extrapolated to the fragments number in experiment. It is seen that, model depicts not badly the experimental angular distributions only for protons. The data for $^2$H, $^3$H, $^3$He differ in experiment from those predicted by SMM. As it is seen from fig.4, there is an explicit trend of cross section increasing at the minimal and maximal ejection angles of the two- and three-nucleon fragments angular dependences in comparison with the model data. As it is known, it may point to the existence of the angular momentum of the fragmenting residual nucleus [15].

## 4. Formation of the helium nuclei
### 4.1. The isotopic composition of the two-charge fragments

Let us see the isotope composition of two-charge fragments in different topological channels (see table 6). Fragments with 10.8<p<16 GeV/c were taken as $^4$He nuclei, and those with p>16 GeV/c as $^6$He ones.

The presented data allow us to mention the following singularities of the helium nuclei formation:

a) Model increases substantially the probability of $^3$He nucleus formation and lessens that of $^4$He one;

b) Probability of $^4$He ($\alpha$-particles) nuclei yield in the channels with only two-charge fragments formation increases as their number arises. The same is also observed in the topological channels, in which along with helium nucleus the two- or four-charge fragments are formed;

c) $^6$He- heavy isotope of helium nucleus is observed only in (2) and (22) topologies (the detailed analysis of $^6$He formation is presented below).

The yields of the helium isotopes in the fragmentation channels (%)

| Part. | W($^3$He) | W($^4$He) | W($^6$He) | ΔW | W($^3$He) | W($^4$He) | W($^6$He) | ΔW |
|---|---|---|---|---|---|---|---|---|
| Topol. | **Experiment** | | | | **SMM** | | | |
| **(2)** | 28.9 | 69.6 | 1.5 | 2.2 | 28.7 | 58.9 | 12.4 | 1.9 |
| **(22)** | 21.9 | 77.4 | 0.7 | 1.1 | 30.8 | 65.9 | 3.3 | 1.0 |
| **(222)** | 15.2 | 84.8 | _ | 0.8 | 22.7 | 76.5 | 0.8 | 0.9 |
| **(2222)** | 13.7 | 86.3 | _ | 1.9 | 55.3 | 44.7 | _ | 5.7 |
| **(23)** | 19.5 | 80.5 | _ | 2.7 | 38.3 | 57.4 | 4.3 | 1.8 |
| **(223)** | 11.8 | 88.2 | _ | 2.7 | 21.3 | 78.7 | _ | 1.9 |
| **(224)** | 13.9 | 86.1 | _ | 5.4 | 52.9 | 47.1 | _ | 4.2 |
| **(24)** | 17.7 | 82.3 | _ | 3.1 | 33.1 | 65.4 | 1.5 | 1.4 |
| **(25)** | 21.0 | 79.0 | _ | 3.3 | 23.0 | 77.0 | _ | 1.2 |
| **(26)** | 15.80 | 84.2 | _ | 2.5 | 30.6 | 69.4 | _ | 4.0 |

### 4.2. Formation of $^4$He nuclei

In dependence of the excitation extent, the fragmentation process may proceed via short-lived unstable fragments formation or that of excited resonance states of the initial nucleus. In connection with it, the search of the intermediate states of the multi-nucleon systems fragmenting on α-particles was conducted.

Let us see the angular distribution of α-particles by all possible topologies of their formation, which is presented in fig.5. Let us mark that, the maximum location is not changed with the topology, but distribution width depends on the extent of the nucleus decay; at $\sum Z_f \geq 7$ it is 1.5 times as tight as at $\sum Z_f < 7$. The maximal value of distributions and the mean ejection angle of α-particles by the model is slightly more: $<\theta>_{exp} = (0.82 \pm 0.02)^0$ and $<\theta>_{smm} = (1.02 \pm 0.01)^0$.

Let us observe the angular correlations between α-particles, which may give some information of their formation mechanisms. The distribution by the angles between α-particles is presented in fig.6 (the solid hystogramm). The substantial singularity of this spectrum is the irregular structure, i.e. existence of two peaks, one of them is absent in the background distribution (the solid curve). The background was obtained for random events made up by mixture of α-particles from different events by taking into account of their multiplicities. The extrapolation of background and experimental distributions is fulfilled in the $\theta_{\alpha\alpha}>1$ region, where as it is seen from fig.6, is observed the good agreement.

It must be pointed that, such a structure is absent in angular distributions between α-particles and fragments, i.e. there isn't peak at the angles near to zero degree.

The tight peak at the minimal relative angles has a $\sigma(\theta_{\alpha\alpha})<0.2^0$ width, which lies within the $\theta_{\alpha\alpha}$ measurement error, and therefore the true width of this peak is less than above mentioned value. It's natural to assume that, if there is a correlated yield of the pair of α-particles, then their relative momenta will be little. The most probable process causing the formation of such pairs may be the decay of the slightly excited system on two $^4$He, for example, that of the unstable $^8$Be nucleus. Stepping of the extreme value of the angle between the correlated α-particles at $\theta_{\alpha\alpha}=0.2^0$, one may estimate the maximal ejected energy-E* of the slightly excited system decay. It's obvious, that the $\theta_{\alpha\alpha}$ will have the maximal value in laboratory frame if decaying α-particles in their center mass system are ejecting transversely to the decaying system movement direction. Then $E^*=P_\perp^2/M\alpha$, where $P_\perp$ can be determined through the extreme angle $\theta_{\alpha\alpha}=0.2^0$ and the mean value of α-particle momentum-$<P\alpha>$ in laboratory frame:

$$P_\perp \approx <P\alpha>\sin(\theta_{\alpha\alpha}/2)=22.6 \text{ MeV/c.}$$

The obtained from such a method value E*=0.15 MeV is the nearest to the full kinetic energy transferred to α-particles, formed at decaying of the unstable $^8$Be nucleus in its ground state $(0^+)$, which equals to 0.1 MeV. The other states of $^8$Be nucleus, such as first $(2^+)$ and second $(4^+)$ ones have higher excitation energies (3.04 and 11.4 MeV respectively) with respect to that of the ground state, and can't cause the appearance of the tight angular correlations between formed $\alpha$-particles [16].

The unstable $^9B$ nucleus with E*=0.3 MeV energy release in its ground state may be either the possible source of formation of the pair of $\alpha$-particles with the relatively small ejection angle. Actually by SMM, in which the formation of the unstable fragments decaying on the $\alpha$-particles ($^8Be \rightarrow \alpha + \alpha$ and $^9B \rightarrow \alpha + \alpha + p$) is taking into account, as it is seen from fig.6(the dashed hystogramm with relative scale), there are also the tight correlations at the same angle values as in the experiment. We must point that, the tight correlations in yield of two $\alpha$-particles and proton, expected in case of an unstable $^9B$ nucleus formation, aren't derived.

So we may conclude that, the observed tight correlations are mainly caused by the intermediate unstable $^8Be$ nucleus decay in its ground state.

An excess of the experimental spectrum over the background in $\theta_{\alpha\alpha}$ =(0.5÷1)$^0$ angular range may be explained by the $\alpha$-particles formation via the first excited $^8Be$ nucleus state decay.

We must point that observed pair correlations in $\alpha$-particles yield in small relative angles may be partly caused by effects of similarity and final state interactions [17].

The unstable nuclei, such as $^5He$ and $^5Li$ formed at the oxygen nucleus fragmentation process in its ground and excited states, may be the supplementary source of $\alpha$-particles formation. Although the binding energy of these nuclei exceeds of 5 MeV/nucleon, we showed that it was sufficient for the decay by the channels either with two $\alpha$-particles formation in case of $^8Be$ or $\alpha$-particle and one nucleon in that of $^5Li$ nuclei. The present experimental data [18] allowed us to estimate the cross section of an $^5Li$-unstable isotope formation. The corresponding results are presented below. As $^5Li$ nucleus decays on $\alpha$-particle and proton in its ground and excited states, the search of formation of that one has been performed by the correlation analysis of its decay products yield.

The events having at least one $^4$He nucleus and proton were selected from total statistic for further analysis. It is impossible the individual identification of the $^5$Li→α+p decay channels, because the nucleus fragments have mainly restricted transferred transverse momentum, and the little ejection angle in laboratory frame respectively. Therefore the yield of $^5$Li nuclei has been studied by comparison of the angular distributions between α-particle and proton ($\Theta\alpha p$) with the background one, obtained by taking into account the separate topological channels part. The background distribution was built up for the random events made up of α-particle from one event and proton from another. The explicitly kinematical effects causing the azimuthal angular correlations are possible during the residual excited nucleus decay. Preliminary the transverse projections of α-particle and proton momenta were defined versus to the total transverse momentum of fragments in each experimental event.

The obtained results are presented in fig.7. The background distribution (solid curve) was normalized on a number of events at $\Theta\alpha p > 2^0$ angles. It's seen that background depicts well the experiment at great angles (the dark circles), but it has substantial excess of the experimental spectrum over the background one, i.e. where αp-correlations are expected in case of $^5$Li formation in its main state. Actually, the average energy E*≈1.97 MeV is released at the $^5$Li→α+p nucleus decay in its main state. In the rest frame of $^5$Li nucleus the products momenta will equal to 54.4 MeV/c.

It's obvious that the maximal angle $\Theta\alpha p$ between α-particle and proton is realized in laboratory frame only when they are ejected transversely to the $^5$Li movement direction in its rest frame. This angle is $\Theta\alpha p \approx 1.2^0$.

The $^5$Li nuclei formation cross section appeared to be $\sigma(^5\text{Li})=8.4\pm0.5$ mb, that didn't differ strongly in the excitation function value obtained previously by us for the stable isotopes:

$\sigma(^6\text{Li})=12.0\pm1.1$ mb and $\sigma(^7\text{Li})=9.6\pm1.0$ mb **[19].**

Such a substantial yield of an unstable $^5$Li nucleus may be the important argument to the favor of its formation mechanism through the forming of finite multinucleon products in $\alpha$-cluster fragmentation of initial oxygen nucleus.

### 4.3. Formation of $^6$He nuclei

13 events with one two-charge fragment in each with mass number $A_f$=6, and the momentum greater than 16.5 GeV/c were found. The mean value of such fragments momentum equaled to (19.6±0.9) GeV/c is in a good agreement with the expected that of He$^6$ nucleus. The cross section of this heavy helium nucleus isotope yield proved to be (0.60±0.17) mb.

Let us see the correlations of fragments and particles formation in the topological channels with the $^6$He-nucleus yield. The fragments with the charges three and more are absent in these channels, but one supplementary two-charge particle, six of which are $^4$He nuclei, presents in seven events.

The mean values of the ejection angle $<\theta>$ and transverse momentum $<P\perp>$ of the helium nucleus isotopes in the topological channels with the number of two-charge particles n($Z_f$=2)=1 and 2 are presented in tab.7.

**Table 7**

The mean values of the ejection angle-$<\theta>$ and the transversal momentum-$<P\perp>$ of the helium nuclei isotopes in the topological channels with the two-charge nuclei number n($Z_f$=2)=1 and 2.

| Type of fragment | $<\theta>$, in angular minutes | $P\perp$, MeV/c |
|---|---|---|
| $^3$He | 84 ± 2 | 234 ± 7 |
| $^4$He | 49 ± 1 | 185 ± 3 |
| $^6$He | 52± 8 | 296 ± 45 |
| $^6$Li | 44 ± 2 | 260 ± 16 |

In the same place those values of $^6$He nucleus are presented for comparison either. As it is seen from table, $<\theta>$ value of $^6$He nucleus isn't differed within the errors with those of $^4$He and $^6$Li ones. It may point that the

«coalescence» of α-cluster with two neutrons during the fragmentation process of the excited nucleus might be the possible mechanism of the formation of nuclei with mass number $A_f=6$.

### 5.Breakup on multinucleon fragments

Let us see the relativistic oxygen nuclei breakup channels on two and more multinucleon fragments with a total nucleon number $\Sigma A_f=16$.

The data on cross sections of the oxygen nucleus corresponding fragmentation channels are presented in tab.8. The corresponding data of SMM are shown in the same table either. Let us remind that formation of the unstable intermediate $^9B$ and $^8Be$ has been taking into account in the used version of the model. It's seen from table that, the channels with the even-even nuclei formation dominate at the oxygen nucleus breakup on multinucleon fragments, herewith, $^4He$ nuclei give the most contribution to the cross section. The part of the multinucleon events output with conservation of all nucleons in the (22) and (2222) (the charges of fragments are pointed out in brackets) topological channels accounts for about 2/3. Let's point that the minimal excitation energy of oxygen nuclei is needed only for formation of the above mentioned topologies (7.41 and 14.4 Mev respectively).

### Table 8
The cross sections of the oxygen nucleus fragmentation on two and more multinucleon fragments.

| Breakup Channels | Cross section, in mb | |
|---|---|---|
| | Experiment | SMM |
| $^{12}C\,^4He$ | $6.61 \pm 0.66$ | $2.49 \pm 0.19$ |
| $^4He\,^4He\,^4He\,^4He$ | $2.10 \pm 0.38$ | $0.015 \pm 0.015$ |
| $^{14}N\,^2H$ | $1.47 \pm 0.29$ | $0.62 \pm 0.10$ |
| $^6Li\,^4He\,^4He\,^2H$ | $0.27 \pm 0.12$ | $0.015 \pm 0.015$ |
| $^{10}B\,^4He\,^2H$ | $0.16 \pm 0.10$ | $0.015 \pm 0.015$ |

Let us see further the data of the oxygen nucleus fragmentation on multicharge fragments with the total charge equaled to the initial nucleus charge. Such breakup channels are realized at the little transfers in the peripheral collisions and as the consequence of it must be sensitive to the initial nucleus structure and the slightly excited fragmenting nucleus mechanisms of nucleons clustering.

Besides of mentioned in table, one candidate in each $^{13}C^{13}H$, $^{12}C^2H^2H$, $^{11}C^3H^1H$ channels was found.

The main features of observed topological channels with $\Sigma Z_f=8$, i.e. (224), (26), (2222), where $\sigma$-topological cross section in mb, Nev-number of events, $<n_{ch}>$, $<n^->$ and $<n_p>$ - the mean multiplicities of one-charge particles, negative pianos and protons, respectively. $W(^4He)$ and $W(^{12}C)$-probabilities of the even-even helium and carbon nuclei isotopes yield, $W(\Sigma A_f=16)$ – the part of events with conservation of all projectile nucleus nucleons in the multicharge fragments, are presented in tab.9.

As it is seen from tab.9, an isotopic composition of two-charge fragments, which main part is $\alpha$-particles (>84%), proved to be the same within the statistical errors. Contrary to that, in model this composition depends on topology and the part of $^3He$ nuclei is significantly more in comparison with that of experiment.

An analysis of momentum spectrums of fragments with $z_f=4$ and 6 showed that:

a) Analysis of four-charge fragments composition showed that they consisted of $^7Be$ nuclei as in model, as in experiment. Consequently, the total number of neutrons, contained in multicharge fragments, is always less than that in the initial projectile-nucleus;

b) The main part of six-charge fragments is $C^{12}$ nucleus ($\approx 70$ %), but in the model the overwhelming part of formed carbon nucleus consists of lighter isotopes ($\approx 83$%).

Above obtained isotopic effects must be probably reflected on the data of one-charge particles multiplicity. Formation of $\pi^-$ mesons at conservation of the initial nucleus charge in secondary fragments takes place most probably in the inelastic collisions processes of target-proton with one or some of projectile-nucleus neutrons, which must cause the decreasing of their number in observed multicharge fragments as compared with protons. We may expect by this cause that in (224) channel (containing always one or more neutron-insufficient nuclei) the negative pianos mean multiplicities, and consequently those of one-charge particles must be greater than in others, as it is observed in experiment. An existence of such a correlation kind is also seen from SMM calculated data.

**Table 9**

The characteristics of the fragmentation channels with $\Sigma Z_f=8$

| Topology | | σ, mb | $<n_{ch}>$ | $<n->$ | $<n_p>$ | W($^4$He) | W(16 ) |
|---|---|---|---|---|---|---|---|
| **(2222)** | **Exp.** | 3.44±0.34 | 1.26±0.10 | 0.13±0.07 | 0.77±0.05 | 86.5±6.5 | 64±11 |
| | **SMM** | 0.29±0.07 | 2.26±0.25 | 0.63±0.13 | 0.26±0.15 | 44.7±7.6 | 5±5 |
| **(224)** | **Exp.** | 0.91±0.17 | 1.78±0.22 | 0.39±0.11 | 0.70±0.11 | >85 | 0 |
| | **SMM** | 1.06±0.13 | 2.31±0.13 | 0.66±0.09 | 0.44±0.09 | 47.1±5.8 | 0 |
| **(26)** | **Exp.** | 10.15±0.6 | 1.25±0.05 | 0.13±0.01 | 0.77±0.05 | 84.4±2.5 | 65±6 |
| | **SMM** | 23.3±0.6 | 1.80±0.03 | 0.40±0.02 | 0.55±0.02 | 69.5±1.2 | 11±1 |

In two other similar by many features (2222) and (26) channels, the fragments formation process takes place with the initial nucleus nucleons conservation in about of 2/3 events, when the part of such ones accounts for approximately 10% in model. In this events the fragments consist only either of α-particles or of α-particle and $^{12}$C. For separation of channels with $^{12}$C ˉnucleus yield the 36.5-42 GeV/c momentum interval has been chosen.

Let us see in details the most statistically provided breakup channel on multicharge fragments- (26) topology.

The (26) channel cross section with respect to the $n_{ch}$-number of one-charge particles is presented in table 10. It's seen that these fragments are accompanied with only by one positive particle yield in overwhelming part of studied channel, herewith, the protons part with P<1.2 GeV/c momentum accounts for nearly 80%.

**Table 10**

The (26) channel cross-section with respect to the $n_{ch}$-number of one-charge particles.

| $n_{ch1}$ | 1 | 3 | 5 |
|---|---|---|---|
| σ, in mb. | 8.94±0.62 | 1.12±0.20 | 0.09±0.06 |

Analysis of momentum spectrums showed that, $^4$He and $^{12}$C nuclei were formed mainly in events with $n_{ch}$=1 and proton yield (their cross section accounts for (6.82±0.51) mb). The separation of fragments by mass was conducted with respect to their momentum value: fragments with z=2 charge in P=(10.8÷15.0) GeV/c momentum interval were referred to α-particle, and those with z=6 and P= (36.5÷44.0) GeV/c to $^{12}$C nuclei.

We further observed the interactions with $P_p$<0.5 GeV/c and $\theta_p$>70, which might be referred to the following reaction:

$$^{16}O+p \rightarrow p+^4He+^{12}C \quad (1).$$

Distribution by proton and α-particle azimuthal angles difference Δφ = $|\varphi_p - \varphi_\alpha|$ is strongly asymmetric with significant maximum at Δφ>150 (fig.8). As it is about of azimuthal correlations in the proton and carbon nuclei yield, those are not much expressed.

The oxygen nucleus breakup on $^4$He and $^{12}$C nuclei one may consider as result of:

1) quasielastic knockout of oxygen nucleus α-cluster by proton;

2) diffractive excitation of oxygen nucleus with its following decay on observed fragments;

The existence at high energies of this special interaction type - an excitation of collective type nucleus with the further breakup on fragments was predicted theoreticaly [20] (see also [21-23]). The latter process may either proceed through the formation of oxygen nucleus excited states with following energy levels (MeV): 11.26(0$^+$), 11.6(3$^-$), 14.02(0$^+$).., 16.8(4$^+$) and greater (the even and spin states are pointed in brackets) with the resonance state width greater of 0.1 MeV [16].

The obtained results show that, the momentum and angular distributions of protons in this events have a maximum at the little Pp≈225 MeV/c momentum and $\theta_p$≈81$^0$ ejection angle in laboratory frame, which is an evidence that the oxygen nucleus breakup process on two-$^4$He and $^{12}$C fragments has an "elastic" diffractive character. It is a proton scattering on the multinucleon system.

With the test aim of these assumptions the calculations were fulfilled in the frame of following phenomenological model. An intermediate excited state of initial nucleus was formed at the inelastic oxygen scattering on proton. Mass of this nucleus was defined by measuring kinematical features of proton:

$$M(^{16}O^*)=M(\alpha^{12}C)=((E_{op}-E_p)^2-(P\alpha_C-P_p)^2)^{1/2}, \quad (2)$$

where $P\alpha_C=P_O-P_p$, $E_{op}$- the total energy of initial particles, $P_o$ and $P_p$- momentum vectors of projectile-oxygen and secondary proton, respectively. The difference of this mass and the sum of $\alpha$ and $^{12}C$ masses is an ejected kinetic energy of breakup:

$$\Delta E=M(\alpha^{12}C)-M\alpha-M^{12}_C \qquad\qquad (3)$$

An experimental distribution by $\Delta E$ has a form to be close to the symmetric one with respect to the mean value $<\Delta E>=(9.1\pm1.0)$ MeV with mean square error 5 MeV.

The calculations were conducted with an assumption that the excited nucleus breakup had been occurred with an isotropy at $(\alpha^{12}C)$ rest frame in the following consequence:

• the kinetic energies and momenta of $\alpha$ and $^{12}C$ fragments were calculated in this system;

• fragments ejection angles with respect to $(\alpha^{12}C)$ system direction and azimuthal ones were initiated randomly by Monte-Carlo method.

• the transfer to laboratory frame was conducted. For taking into account the influences of the momentum and angle measurements errors,$\Delta P$-, $\Delta\alpha$-,$\Delta\beta$-(errors of momentum, depth angle and flat angle measurements, respectively) were initiated randomly by normal law with the experimentally defined dispersions.

Obtained results are presented in fig.8 (dashed and crossed curve). In the same figure SMM calculations are also presented by dashed curve. It's seen that calculations by phenomenological and SMM models are too close to each other, but those are deviating strongly on experimental data in $\Delta\varphi$ $>160^0$ range. An observed substantial trend of proton and $\alpha$-particle ejection at opposite sides may favor strongly to the mechanism of $\alpha$-cluster knocking out from projectile-nucleus.

It infers from presented data that, at oxygen nucleus fragmentation on two fragments with conservation of all initial nucleus nucleons only single (26) channel with even-even nuclei is fulfilled.

### 6.Conclusion

The results of this paper may be briefly formulated as following:

1.Cross sections of topological channels have been measured. The data show that, with the most probability the multifragmentation breakup of oxygen nucleus is performed, and it is mainly with helium nuclei yields. About 80 percent of helium nuclei are $\alpha$-particles. The noticable part of $\alpha$-particles is products of $^5$Li and $^8$Be - unstable short-life nuclei decay. The same probabilities of "mirror" $^3$H, $^3$He and $^7$Li, $^7$Be nuclei yields are observed.

2. It infers from comparison of experimental data on light fragments yield with those of SMM model that, besides fermi-breakup, there are other mechanisms of their formation: "evaporation" processes in yields of slow fragments in oxygen rest frame and processes of coalescnce of cascade nucleons resulting in formation of the significant part of high-energy fragments with A=2,3.

3.At multifragmentation breakup of oxygen nucleus on multinucleon fragments with the conservation of all target nucleons, breakup channels on four $\alpha$-particles and two $^{12}$C и $^4$He nuclei are mainly fulfilled. The data agree satisfactorily with the mechanism of $\alpha$-cluster knocking out and the diffractive excitation in whole.

# FIGURE CAPTIONS

**Fig. 1.**  Distribution of onechareged positive particles by value x=1/$p_f$.

**Fig. 2.**  Invariant structural function of protons as kinetic energy function f(T)− (a); and its relation to calculations on SMM−(b).

**Fig. 3.**  Distribution of invariant structural function as kinetic energy function f(T). ● – Experiment, ○ – SMM.

**Fig. 4.**  Angle distribution of light fragments in antilaboratory system.

**Fig. 5.**  Angle distribution of α-particles.

**Fig. 6.**  Distribution on angles between α-particles.

**Fig. 7.**  Distribution on yield angles between α-particle and proton.

**Fig. 8.**  Distribution on difference of protons azimutal angles and α-particles.

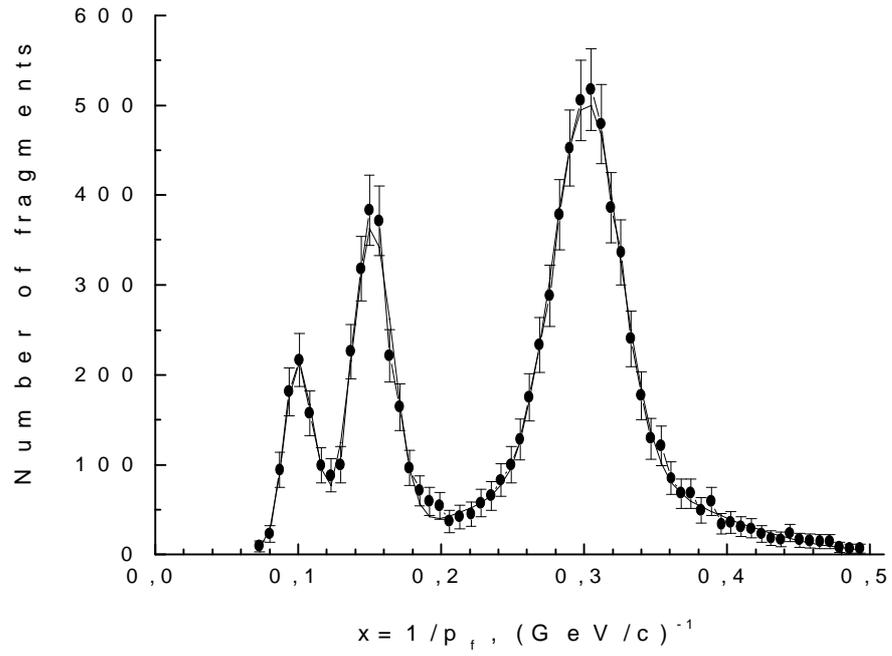

x = 1 / p$_f$ , ( G e V /c )$^{-1}$

F ig . 1

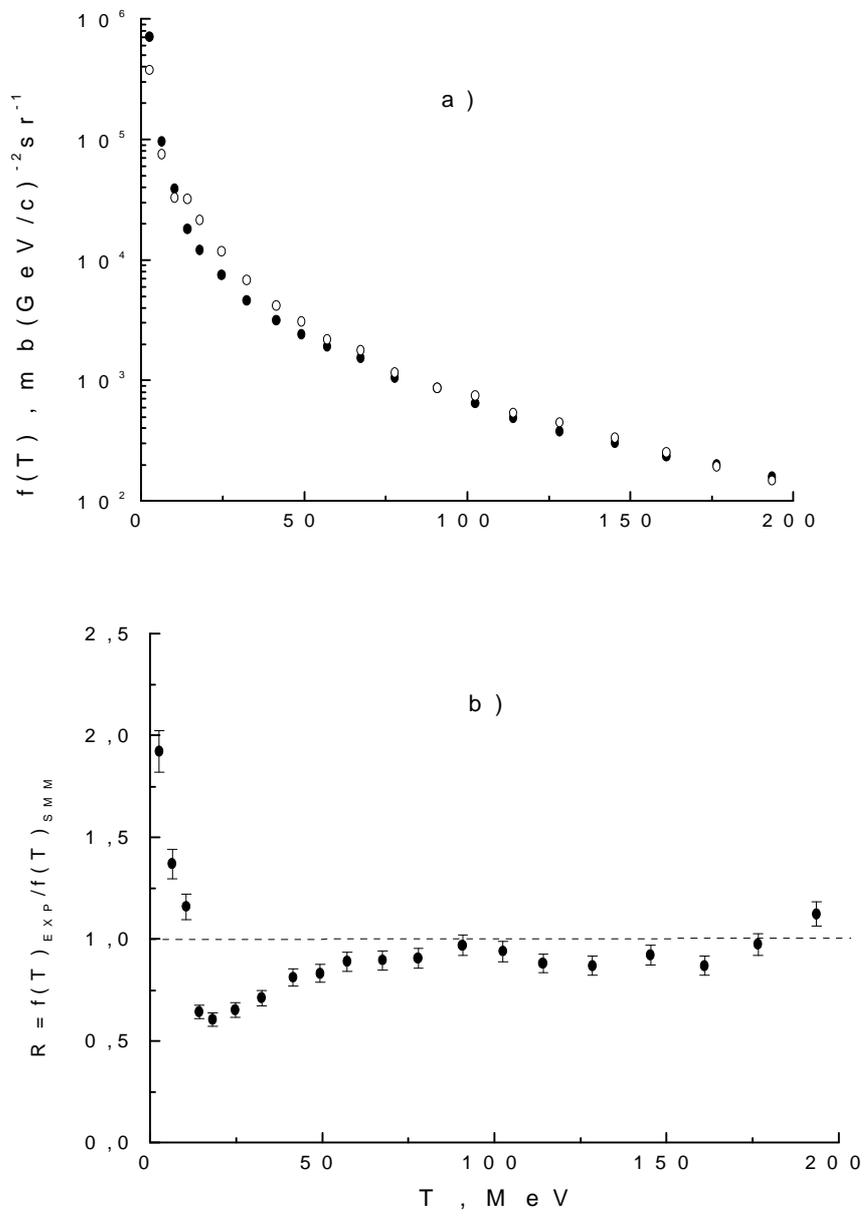

F i g . 2

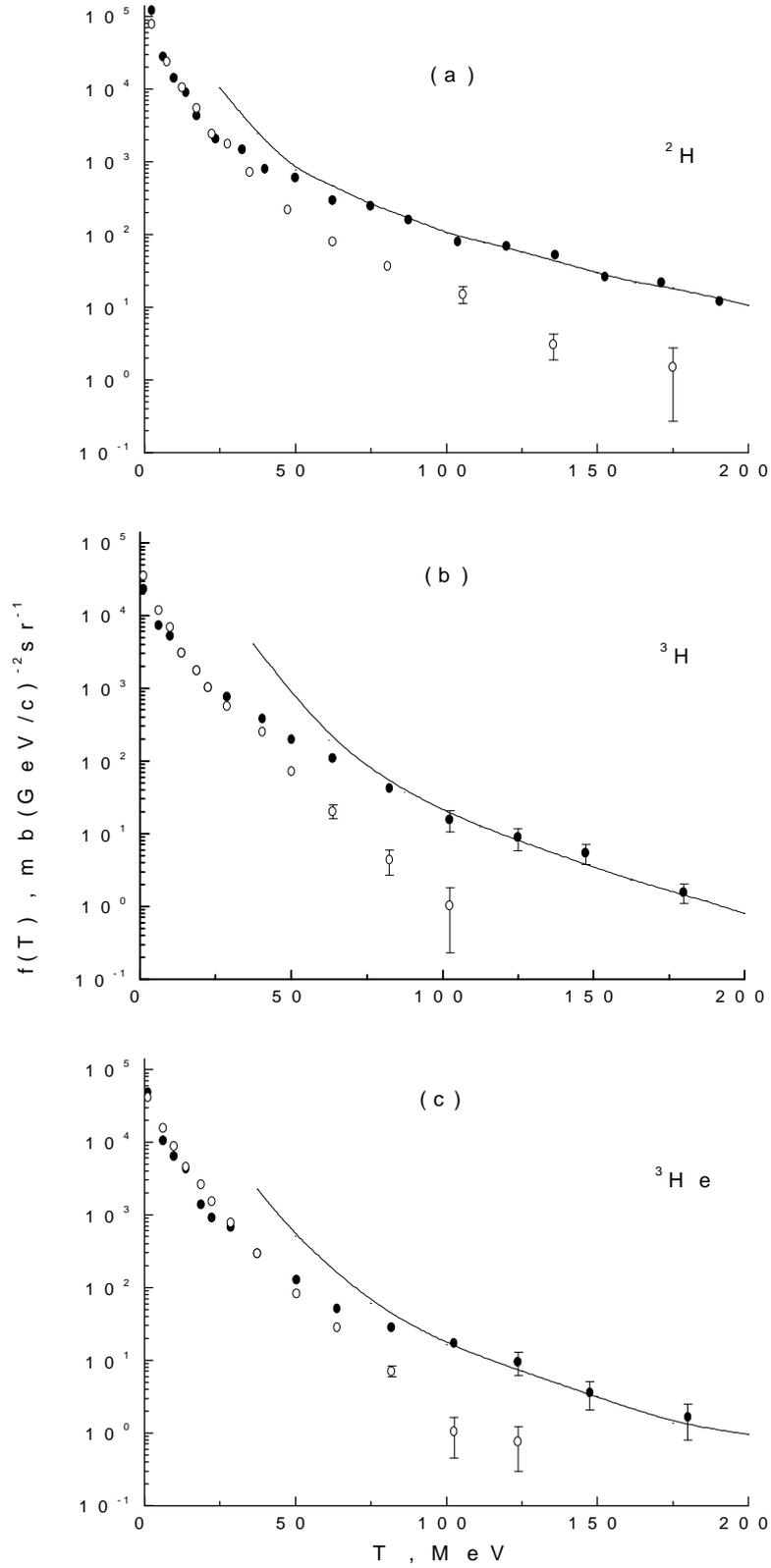

Fig. 3

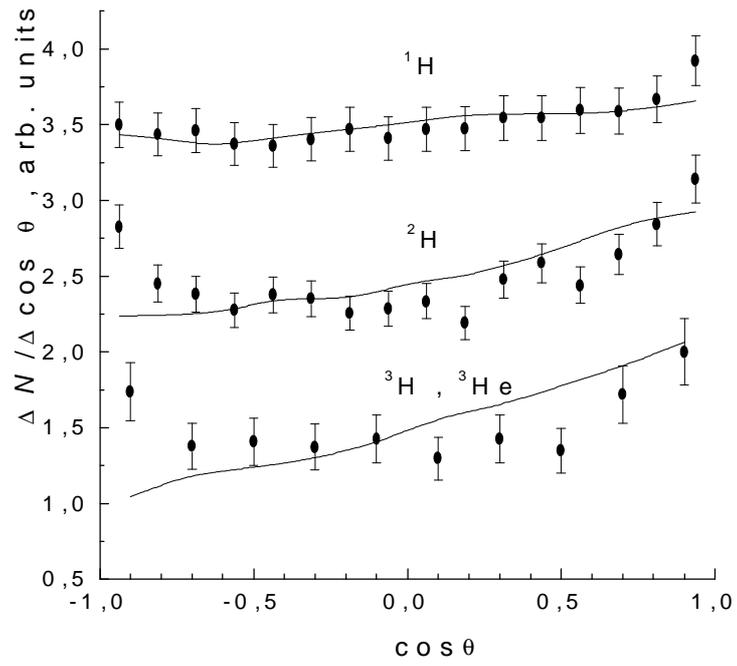

Fig. 4

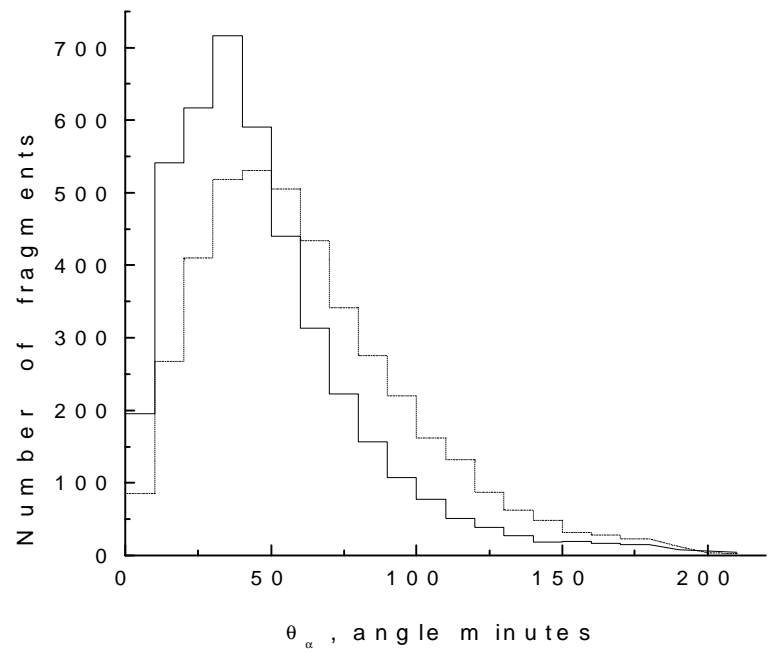

F i g . 5

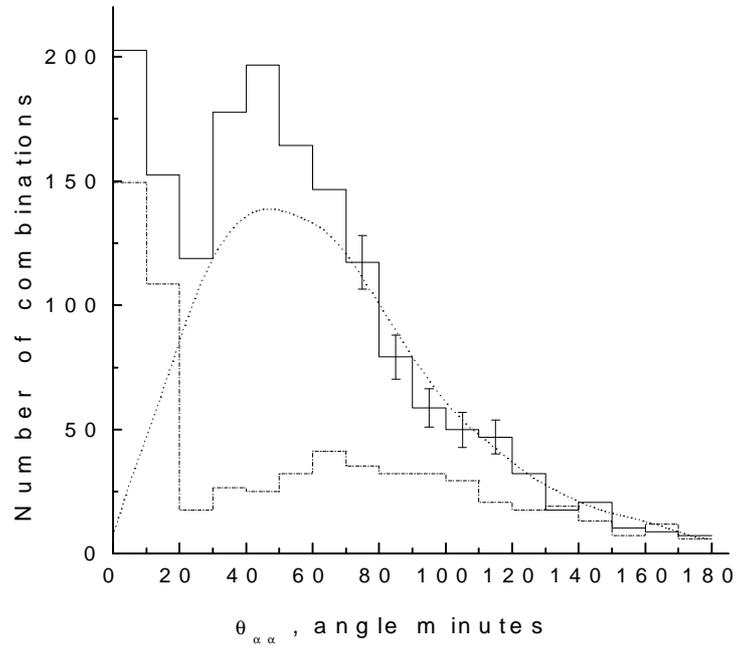

Fig. 6

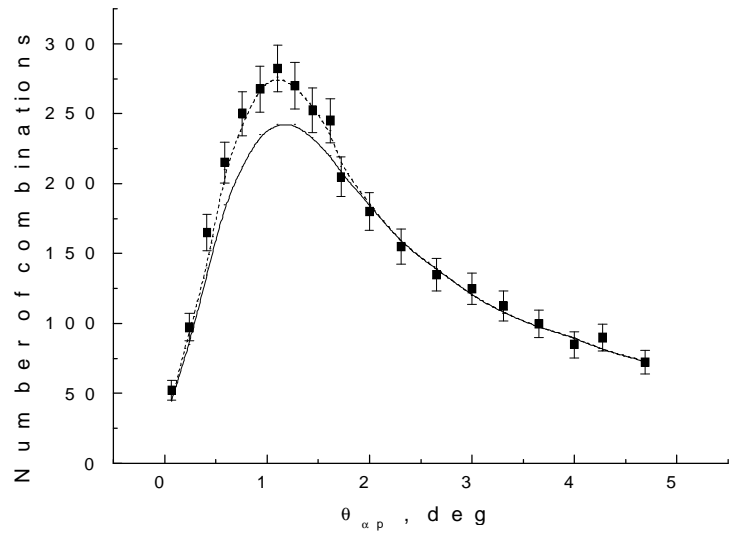

F i g . 7

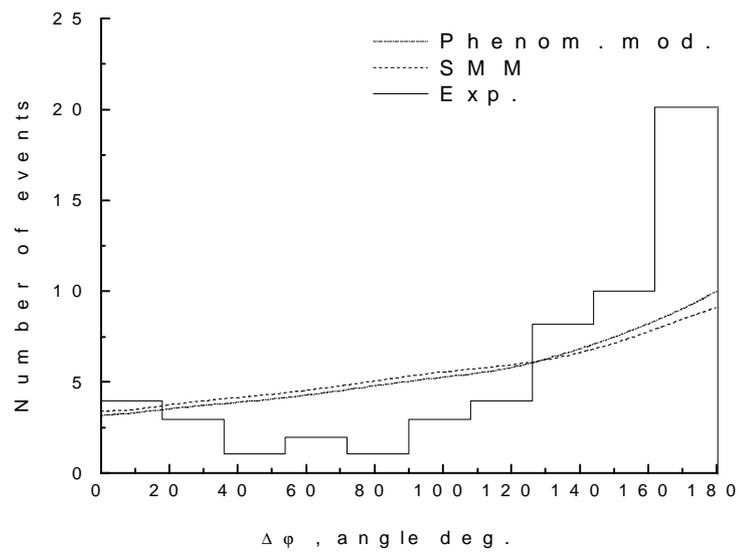

F i g . 8